\documentclass[11pt,english,preprint,aps,prd,showpacs,superscriptaddress,nofootinbib,tightenlines]{revtex4}
\usepackage{babel}
\usepackage{array}
\usepackage{multirow}
\usepackage{amsmath}
\usepackage{amssymb}
\usepackage{graphicx}
\usepackage{amsfonts}
\usepackage{multirow}
\usepackage{mathrsfs}
\usepackage{epstopdf}
\usepackage{bm}
\usepackage{bbm}
\usepackage{color}
\usepackage{array}
\usepackage{diagbox}
\usepackage{float}
\usepackage{braket}
\usepackage{upgreek}
\usepackage{subcaption}
\usepackage{diagbox}
\usepackage{slashed}
\usepackage[figureright]{rotating}

\makeatletter

\@ifundefined{textcolor}{}
{%
 \definecolor{BLACK}{gray}{0}
 \definecolor{WHITE}{gray}{1}
 \definecolor{RED}{rgb}{1,0,0}
 \definecolor{GREEN}{rgb}{0,1,0}
 \definecolor{BLUE}{rgb}{0,0,1}
 \definecolor{CYAN}{cmyk}{1,0,0,0}
 \definecolor{MAGENTA}{cmyk}{0,1,0,0}
 \definecolor{YELLOW}{cmyk}{0,0,1,0}
}

\usepackage{babel}

\@ifundefined{textcolor}{}{%
 \definecolor{BLACK}{gray}{0}
 \definecolor{WHITE}{gray}{1}
 \definecolor{RED}{rgb}{1,0,0}
 \definecolor{GREEN}{rgb}{0,1,0}
 \definecolor{BLUE}{rgb}{0,0,1}
 \definecolor{CYAN}{cmyk}{1,0,0,0}
 \definecolor{MAGENTA}{cmyk}{0,1,0,0}
 \definecolor{YELLOW}{cmyk}{0,0,1,0}
}

\def\OMIT#1{}

\def\hlinew#1{%
  \noalign{\ifnum0=`}\fi\hrule \@height #1 \futurelet
   \reserved@a\@xhline}
\usepackage{array}
\newcommand{\PreserveBackslash}[1]{\let\temp=\\#1\let\\=\temp}
\newcolumntype{C}[1]{>{\PreserveBackslash\centering}p{#1}}
\newcolumntype{R}[1]{>{\PreserveBackslash\raggedleft}p{#1}}
\newcolumntype{L}[1]{>{\PreserveBackslash\raggedright}p{#1}}

\newcommand{\beq}{\begin{equation}}
\newcommand{\eeq}{\end{equation}}
\newcommand{\bqa}{\begin{eqnarray}}
\newcommand{\eqa}{\end{eqnarray}}
\newcommand{\bseq}{\begin{subequations}}
\newcommand{\eseq}{\end{subequations}}

\newcommand{\fverb}{\setbox\fverbbox=\hbox\bgroup\verb}
\newcommand{\fverbdo}{\egroup\medskip\noindent%
\fbox{\unhbox\fverbbox}\ }
\newcommand{\fverbit}{\egroup\item[\fbox{\unhbox\fverbbox}]}
\newbox\fverbbox

\newcommand{\Rmnum}[1]{\expandafter\@slowromancap\romannumeral #1@}

\usepackage{tikz}
\usetikzlibrary {shapes.misc}
\usetikzlibrary{arrows.meta}
\usetikzlibrary{calc}
\usetikzlibrary{
  decorations.pathmorphing,
  decorations.pathreplacing,
  decorations.markings
}
\usetikzlibrary{patterns}

\tikzset{
  every picture/.style={semithick, line cap=round},
  scalar/.style={dashed},
  fermion/.default=0.5,
  fermion/.style={postaction={decorate, decoration={
    markings,
    mark=at position #1 with {\arrow{Stealth[angle=30:7pt,inset=1.5pt]}},
    transform={xshift={3.5pt*cos(15)}}
  }}},
  antifermion/.default=0.5,
  antifermion/.style={postaction={decorate, decoration={
    markings,
    mark=at position #1 with {\arrowreversed{Stealth[angle=30:7pt,inset=1.5pt]}},
    transform={xshift={-3.5pt*cos(15)}}
  }}},
  gluon/.default=3pt,
  gluon/.style={decorate, decoration={
    coil,
    amplitude=0.5*#1,
    aspect=1,
    segment length=#1
  }},
  rgluon/.default=3pt,
  rgluon/.style={decorate, decoration={
    coil,
    amplitude=-0.5*#1,
    aspect=-1,
    segment length=#1
  }},
  gluonpre/.default=0pt,
  gluonpre/.style={decorate, decoration={
    coil,
    amplitude=1.5pt,
    aspect=1,
    segment length=3pt,
    pre length=#1
  }},
  rgluonpre/.default=0pt,
  rgluonpre/.style={decorate, decoration={
    coil,
    amplitude=-1.5pt,
    aspect=-1,
    segment length=3pt,
    pre length=#1
  }},
  crossmark/.style={cross out, draw=red, inner sep=2pt},
  counter/.style={path picture={
    \draw (path picture bounding box.south east) --
      (path picture bounding box.north west)
      (path picture bounding box.south west) --
      (path picture bounding box.north east);
  }},
  cnode/.default=8pt,
  cnode/.style={inner sep=0pt, minimum size=#1, circle}
}

\makeatother
\begin{document}
\title{Branching fraction of $\Xi_{bc}^+\to \Xi_{c}^+ J/\psi$ in the final-state-interaction approach}
\author{Xiao-Hui Hu}

\affiliation{The College of Materials and Physics, China University of mining
and technology, Xuzhou 221116, China \vspace{0.2cm}
 }
 \affiliation{Lanzhou Center for Theoretical Physics, Key Laboratory of Theoretical Physics of Gansu Province, Key Laboratory of Quantum Theory and Applications of MoE, Gansu Provincial Research Center for Basic Disciplines of Quantum Physics, Lanzhou University, Lanzhou 730000, China \vspace{0.2cm}}

\author{Cai-Ping Jia}

\thanks{jiacp@pku.edu.cn, corresponding author}

\affiliation{Center for High Energy Physics, Peking University, Beijing 100871,
China\vspace{0.2cm}
 }
  
\author{Ye Xing}

\affiliation{The College of Materials and Physics, China University of mining
and technology, Xuzhou 221116, China \vspace{0.2cm}
 }

 \author{Fu-Sheng Yu}

\affiliation{Frontiers Science Center for Rare Isotopes, and School of Nuclear
Science and Technology, Lanzhou University, Lanzhou 730000, China\vspace{0.2cm}
 }
\date{\today}
\begin{abstract}
The process of $\Xi_{bc}^{+}\to \Xi_{c}^{+}J/\psi$ is among the most favored modes for searching for bottom-charm baryons. 
However, its branching fraction has never been studied in theory. 
In this work, we investigate the branching fraction of $\Xi_{bc}^{+}\to \Xi_{c}^{+}J/\psi$ in the final-state-interaction approach, as it is dominated by the color-suppressed nonfactorizable contributions. 
A similar process, $\Lambda_{b}^{0}\to \Lambda^0 J/\psi$,
is used as a control mode to fix the model parameter. Consequently,  the branching fraction of 
$\Xi_{bc}^{+}\to \Xi_{c}^{+}J/\psi$ is predicted to be $(1.55_{-0.42}^{+0.50})\times10^{-4}$.
With the production rate of bottom-charm baryons and the detection efficiencies of the final states, it is expected for considerable signal events to observe $\Xi_{bc}^+$ in the near future. 
\end{abstract}

\maketitle
\section{Introduction}
\begin{figure}
	\includegraphics[width=\textwidth]{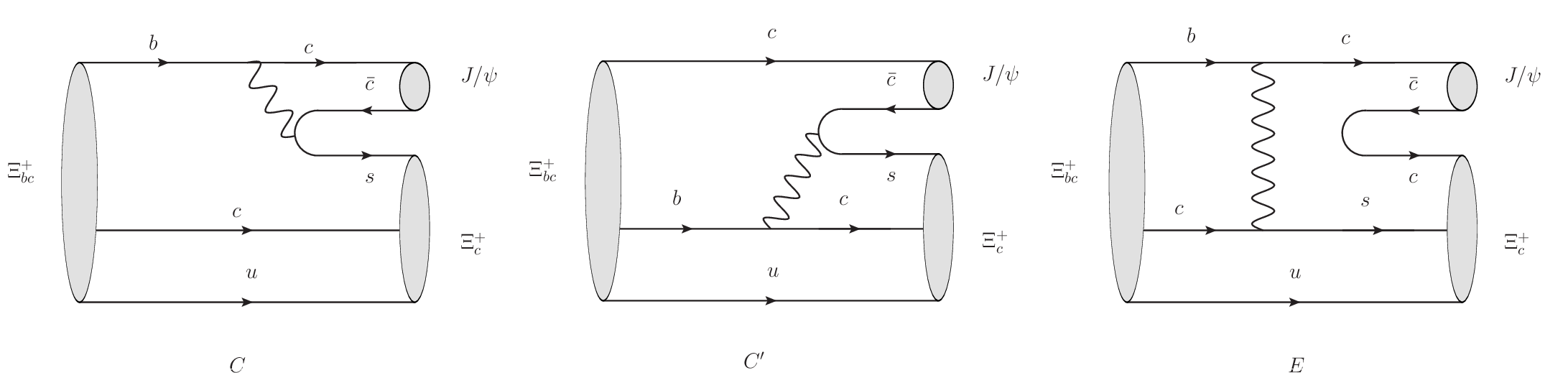}
	\caption{Leading-order Feynman diagram for the decay ${\Xi}_{bc}^{+}\to{\Xi}_{c}^{+}J/\psi$. C: the color-suppressed internal W-emission diagram; E: the W-exchange diagram.}
	\label{fig:tuoputu}
\end{figure}

Doubly heavy baryons have been predicted by quark models for many years. However, only the doubly charmed baryon has been observed, with charmed-bottom and doubly bottom baryons remaining undetected. The LHCb collaboration has searched for charmed-bottom baryons using proton-proton collision data collected from 2016-2018, through the decay channels $\Xi_{bc}^{0}\to D^{0}pK^{-}$ and $\Omega_{bc}^{0}(\Xi_{bc}^{0})\to \Lambda_{c}^{+}\pi^{-}(\Xi_{c}^{+}\pi^{-})$, but no significant signals for the two neutral baryons $\Xi_{bc}^{0}$ and $\Omega_{bc}^{0}$ were observed~\cite{LHCb:2020iko,LHCb:2021xba}.
Subsequently, they performed the first search for the bottom-charmed baryon $\Xi_{bc}^{+}$ using the full proton-proton collision data from Run I and Run II, via the decay channel $\Xi_{bc}^{+} \rightarrow J/\psi\Xi_{c}^{+}$. Local excesses with significances of $4.3\sigma$ and $4.1\sigma$ are observed near masses of 6571~MeV/\(c^{2}\) and 6694~MeV/\(c^{2}\), respectively. After accounting for statistical fluctuations and systematic uncertainties, the global significances are $2.8\sigma$ and $2.4\sigma$~\cite{LHCb:2022fbu}. Compared to the earlier searches for $\Xi_{bc}^{0}$ and $\Omega_{bc}^{0}$, the experimental signature for $\Xi_{bc}^{+}$ is considerably clearer. And the lifetime of $\Xi_{bc}^{+}$ baryon is expected to be longer than that of $\Xi_{bc}^{0}$ and $\Omega_{bc}^{0}$~\cite{Kiselev:1999kh,Kiselev:2001fw,Karliner:2014gca,Berezhnoy:2018bde,Cheng:2019sxr}, leading to higher selection efficiency, because lifetime information helps suppress background from primary $pp$ interactions.
Additionally,  the subsequent decay $J/\psi \to \mu^+\mu^-$ used in this analysis typically has about three times higher selection efficiency compared to fully hadronic modes employed in the previous $\Xi_{bc}^{0}$ and $\Omega_{bc}^{0}$ searches. 
Moreover, regarding the decay of bottom-charmed baryons, if the decay proceeds via the charm quark constituent, favoring decays into strange hadrons, the Cabibbo-Kobayashi-Maskawa (CKM) matrix element $|V_{cs}|$ is the largest. Experimentally, reconstruction is performed step by step through intermediate particles, resulting in a relatively small overall decay branching fraction when considering the entire decay chain. And the c quark decay
channels suffer low reconstruction efficiencies of the b hadrons in their final states~\cite{Han:2021gkl}.
Therefore, current experimental searches focus on decay channels with relatively large total branching fractions and high signal selection efficiencies.
In contrast, decays proceeding via the bottom quark are dominated by the CKM element $|V_{cb}|$, making decays into doubly charmed baryons or two charmed hadrons most probable. While previous searches for $\Xi_{bc}^{0}$ and $\Omega_{bc}^{0}$ relied on suppressed $b \to u$ / $b \to s$ transitions or $W$-exchange mechanisms, the decay $\Xi_{bc}^{+} \to J/\psi\,\Xi_{c}^{+}$ occurs via a color-suppressed $b \to c$ transition, whose amplitude is less susceptible to suppression, as illustrated in Fig.~\ref{fig:tuoputu}.  
Whether this channel indeed represents the optimal discovery channel requires further detailed investigation. Theoretical studies of this specific process remain limited to date. Thus, the present analysis establishes a robust theoretical framework to guide potential experimental discovery of the first charm-bottom baryon.baryon.

As shown in Fig.~\ref{fig:tuoputu}, this decay $\Xi_{bc}^{+}\to\Xi_{c}^{+}J/\psi$, proceeding via the weak $b\to s\bar{c}c$ transition, provides a particularly clean probe into the strong interaction dynamics of heavy hadronic decays~\cite{Neubert:2001sj}. Such decays have been intensively studied in $B\to KJ/\psi$~\cite{CDF:1995izg,Belle:2002oex,BaBar:2004htr,Cheng:2000kt} and $\Lambda_{b}\to \Lambda_{c}J/\psi$ decays.
The decays of $b$ hadrons into $J/\psi$ are experimentally advantageous due to the distinctive signature of the subsequent $J/\psi \rightarrow \mu^+\mu^-$ decay. In this work, ${\Xi}_{bc}^{+}\to{\Xi}_{c}^{+}J/\psi$ will show even richer QCD dynamics than the above two decay channels due to the additional heavy quark. 
As indicated in bottom meson decay~\cite{Cheng:2004ru}, the branching ratios of $\bar{B}\to D^{(*)0}\pi^{0}$ are all significantly larger than theoretical expectations based on naive factorization. This difference can be explained through the long-distance final-state-interaction (FSI) contributions, it induced large magnitude and phase of the ratio between the W-emission tree diagram (T) and the W-emission diagram (C). Moreover, diagram C is dominated by nonfactorizable contributions. This means the nonfactorization contribution of C diagram is considerably important compared to the naive factorization estimation.  
Moreover, the process $\Xi_{bc}^{+}\to\Xi_{c}^{+}J/\psi$ occurs through the C diagram and fully nonfactorized W-exchange diagram, it is expected the contribution of nonfactorization amplitudes can be estimated through the FSIs via a rescattering mechanism.
This process is particularly interesting theoretically as it proceeds exclusively through the color-suppressed internal W-emission diagrams~\cite{Leibovich:2003tw}.
Similar to the analogous $\Lambda_b$ decays~\cite{Mohanta:1998iu,Cheng:1996cs,Fayyazuddin:1998ap,Ivanov:1997hi,Wei:2009np,Ivanov:1997ra,Cheng:1991sn,Cheng:1995fe,Mott:2011cx,Fayyazuddin:2017sxq,Chou:2001bn,Zhu:2018jet,Hsiao:2015cda,Gutsche:2017wag,Gutsche:2013oea,Gutsche:2015lea,Ajaltouni:2004zu,Hsiao:2015txa,Gutsche:2018utw}, significant nonfactorizable contributions are expected in $\Xi_{bc}^{+}\to \Xi_{c}^{+}J/\psi$. These contributions offer crucial information for understanding the nonfactorizable mechanism in heavy baryon decays.

\begin{figure}
	\includegraphics[width=0.45\textwidth]{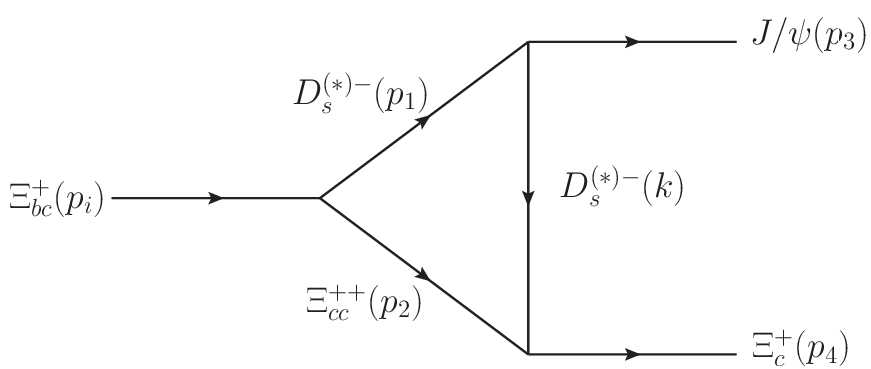}
	\caption{Triangle diagram at the hadron level for the decay ${\Xi}_{bc}^{+}\to{\Xi}_{c}^{+}J/\psi$.}
	\label{fig:triangle}
\end{figure}

This color-suppressed mode receives significant nonfactorizable contributions, challenging the factorization ansatz.
From a theoretical perspective, a key challenge lies in evaluating the nonfactorizable contribution. The FSI approach provides a powerful framework for studying doubly heavy baryon decays~\cite{Hu:2024uia}, having been successfully applied to various $B$ meson and heavy baryon decay processes~\cite{Cheng:2004ru,Jia:2024pyb}.
The basic ingredient is that the short-distance contributions can be factorized into the nonperturbative transition
matrix element $\langle{\Xi}_{bc}|(V-A)_{\mu}|{\Xi}_{c}\rangle$,
characterized by form factors, and the matrix element $\langle V|(V-A)_{\mu}|0\rangle$,
represented by the decay constant of charmonium $J/\psi$, and the computation of nonfactorizable long-distance contributions is achieved via the rescattering mechanism. 
The final states of this nonleptonic weak decay, $\Xi_{c}$ and $J/\psi$, involve their mutual scattering
via the exchange of one particle, resulting in a triangle diagram
at the hadron level, as depicted in Fig.~\ref{fig:triangle}. To
circumvent the issue of double counting, contributions from both short
and long distances are segregated into the tree emission process $\Xi_{bc}^{+}(p_{i})\to\Xi_{cc}^{++}(p_{1})D^{(*)-}_s(p_{2})$
and final state interaction effects. 
Previous literature~\cite{Yu:2017zst,Han:2021azw,Jiang:2018oak,Li:2020qrh,Han:2021gkl}
has utilized the optical theorem and Cutkosky cutting rules to compute
triangle diagrams. However, these calculations only yield the imaginary
component of the amplitude, making it challenging to represent the strong phase accurately. 
We have recently corrected this issue by employing
the comprehensive analytical expression of loop integrals, which allows
us to ascertain both the magnitude and strong phase of triangle diagrams and analyzed the nonleptonic
decays of ${\cal B}_{cc}\rightarrow {\cal B}_{c} P$ by using the FSI approach and obtained satisfactory results.
In this work, we will extend this approach to calculate the complete triangle diagrams and focus on the study of $\Xi_{bc}^{+}\to \Xi_{c}^{+}J/\psi$ decays.
After employing the effective Lagrangian at the hadronic level of the strong vertex in the rescattering mechanism, we can compute the final state interaction effects. 

\begin{figure}
	\includegraphics[width=0.45\textwidth]{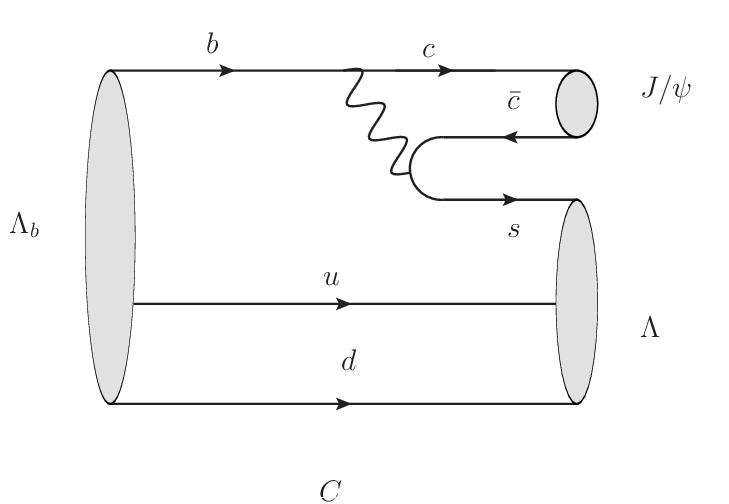}
	\caption{Leading-order Feynman diagram for the decay $\Lambda_b \rightarrow \Lambda J/\psi$.}
	\label{fig:Lambda}
\end{figure}

While in the calculations of the final states interaction effects,
the nonperturbative parameters such as the cut-off $\Lambda$ in
the loop calculation will bring large theoretical uncertainties on
the branching ratios. Unlike the case of B meson~\cite{Ablikim:2002ep,Cheng:2004ru},
there are not enough data to determine the nonperturbative parameters of the charmed-bottom baryon, while the expermental data of the similar bottom decay channel $\Lambda_b \rightarrow \Lambda J/\psi$, as shown in Fig.~\ref{fig:Lambda}, allows us to constrain this model parameter.
The well-studied $\Lambda_b \rightarrow \Lambda J/\psi$ process serves as a benchmark for model calibration, enabling reliable theoretical predictions. The $\Lambda_b \rightarrow \Lambda J/\psi$ decay was first observed by the UA1 Collaboration at the CERN $p\bar{p}$ collider~\cite{UA1:1991vse}, with subsequent detailed studies conducted by CDF~\cite{CDF:1992lrw,CDF:1996rvy,CDF:2006eul} and D0~\cite{D0:2007giz,D0:2004quf,D0:2011pqa,D0:2012hfl} at the Fermilab Tevatron. However, the absolute branching fraction remains undetermined, primarily due to uncertainties in the hadronization fraction of $b$ quarks into $\Lambda_b$ baryons.
The current Particle Data Group (PDG) reports the combined measurement~\cite{ParticleDataGroup:2024cfk}
\begin{equation}\label{eq:bbb}
    f(b\rightarrow\Lambda_b) \times \mathcal{B}(\Lambda_b \rightarrow \Lambda J/\psi) = (5.8 \pm 0.8) \times 10^{-5},
\end{equation}
where $f(b\rightarrow\Lambda_b)$ denotes the fragmentation fraction of $b$ quarks into $\Lambda_b$ baryons.
This observation provides valuable insights for investigating the dynamics of the charmonium decay mode $\Xi_{bc}^{+}\to \Xi_{c}^{+}J/\psi$.
Moreover, nonperturbative form factors, decay constant, and strong coupling constants serve as critical input parameters.
The nonperturbative form factors have been assessed in a variety of theoretical studies,
utilizing the light cone sum rules (LCSR) \cite{Shi:2019fph,Hu:2019bqj,Hu:2022xzu,Aliev:2022tvs,Aliev:2022maw},
light-front quark model (LFQM)~ \cite{Wang:2017mqp,Zhao:2018mrg,Cheng:2020wmk,Ke:2019lcf,Ke:2022gxm,Hu:2020mxk},
QCD sum rules~\cite{Shi:2019hbf}, QCD factorization~\cite{Sharma:2017txj,Gerasimov:2019jwp},
diquark effective theory~\cite{Shi:2020qde}, constituent quark model~\cite{Gutsche:2018msz,Gutsche:2019iac},
and heavy quark effective theory~\cite{Sharma:2017txj,Dhir:2018twm}.
In this work, we adopt the theoretical predictions of form factors
within LFQM \cite{Hu:2020mxk} as inputs. The decay constant of
$J/\psi$ is avilable in PDG~\cite{ParticleDataGroup:2024cfk}.  In our previous work~\cite{Hu:2025ajx}, we have calculated the strong coupling constants using light cone sum rules.

The rest of the paper is organized as follows. In Sec.~\ref{sec:framework}, we give a brief review of the final-state interactions theory framework and define relevant observation for ${\Xi}_{bc}^{+}\to{\Xi}_{c}^{+}J/\psi$ decays. In Sec.~\ref{sec:results}, the numerical analysis including the value of model parameter $\eta$, the result of the branching ratio and the signal events, are given. The summary and conclusions are given in Sec. IV.

\section{Theoretical Framework}
\label{sec:framework}
\subsection{Calculation of short-distance amplitudes under the factorization
hypothesis}

\label{subsec:short}

At the tree level, the nonleptonic weak decay of bottom-charmed baryons
are induced by the decay of the bottom quark. The effective Hamiltonian
can be represented as follows, 
\begin{align}
\mathcal{H}_{eff}=\frac{G_{F}}{\sqrt{2}}V_{cs}^{\ast}V_{bc}[C_{1}(\mu)O_{1}(\mu)+C_{2}(\mu)O_{2}(\mu)]+\text{H}.\text{c}.\label{eq:hamilton}
\end{align}
Here $V_{bc}$ and $V_{cs}$ represent the CKM matrix elements. And the four-fermion operators $O_{1}$ and
$O_{2}$ can be expressed as, 
\begin{align}
O_{1}=(\bar{s}_{\alpha}c_{\alpha})_{V-A}(\bar{c}_{\beta}b_{\beta})_{V-A},
\quad O_{2}=(\bar{s}_{\alpha}c_{\beta})_{V-A}(\bar{c}_{\beta}b_{\alpha})_{V-A},
\end{align}
with the color indices $\alpha$ and $\beta$. And $C_{1,2}(\mu)$
are the relevant Wilson coefficients. After inserting the effective
Hamiltonian as Eq.~(\ref{eq:hamilton}), the amplitude of the nonleptonic
weak decay $\Xi_{bc}^{+}\to\Xi_{c}^{+}J/\psi$
can be evaluated with the hadronic matrix element, 
\begin{align}
\langle\Xi_{c}^{+}J/\psi|\mathcal{H}_{eff}|\Xi_{bc}\rangle=\frac{G_{F}}{\sqrt{2}}V_{cs}^{\ast}V_{bc}\sum_{i=1,2}C_{i}\langle\Xi_{c}^{+}J/\psi|O_{i}|\Xi_{bc}\rangle.\label{eq:hmebc}
\end{align}
At the tree level, this process is dominated by the color-suppressed $C$ diagram, its short-distance
contribution can be given via its relation to the $T$ diagram after
Fierz transformation.
In this work, we neglect the penguin operators for the strong suppression
of CKM matrix elements in the bottom hadron decays. \label{sec:short}
According to the factorization hypothesis, the matrix element $\langle\Xi_{c}^{+}J/\psi|O_{i}|\Xi_{bc}\rangle$
in Eq.~(\ref{eq:hmebc}) can be factorized into two parts. The first
part is parametrized with the decay constant of the emitted meson,
while the second part can be evaluated using the heavy-light transition
form factors. About the processes containing the final state of charmonium $J/\psi$, QCD factorization is still applicable due to the small transverse size of the charmonium in the heavy quark limit~\cite{Cheng:2000kt}. A combined coefficient ${\bar a}_2$ extracted from the experiment data of $B\to J/\psi K$ is $|{\bar a}_2|_{\rm expt}=0.26$ is close to the value of $a_2$ given in Ref.~\cite{Zhu:2018jet}. The short-distance contribution of the $C$ diagram
can be expressed as, 
\begin{align}
\langle\Xi_{c}^{+}J/\psi|\mathcal{H}_{eff}|\Xi_{bc}\rangle_{SD}^{C}&=\frac{G_{F}}{\sqrt{2}}V_{cs}^{\ast}V_{bc}a_{2}(\mu)\langle J/\psi|\bar{c}\gamma^{\mu}(1-\gamma_{5})c|0\rangle\langle\Xi_{c}|\bar{s}\gamma_{\mu}(1-\gamma_{5})b|\Xi_{bc}\rangle,\label{eq:hadronicXi}
\end{align}
here $a_{2}(\mu)=C_{1}(\mu)/3+C_{2}(\mu)$ is the effective Wilson
coefficients. And under the bottom scale $\mu=m_{b}$, the
Wilson coefficients are taken as $C_{1}(\mu)=1.117$ and $C_{2}(\mu)=-0.268$~\cite{Zhu:2018jet}.
The first matrix element in Eq.~(\ref{eq:hadronicXi}) can be defined
using the decay constant of the emitted vector meson $J/\psi$, denoted
by $f_{J/\psi}$, respectively. 
\begin{align}
\langle J/\psi(q)|\bar{c}\gamma^{\mu}(1-\gamma_{5})c|0\rangle & =m_{J/\psi}f_{J/\psi}\varepsilon^{\ast\mu}(q),\label{eq:Vdecay}
\end{align}
where $\varepsilon^{\mu}$ represents the polarization vector of $J/\psi$ and $m_{J/\psi}$ is the mass of vector meson $J/\psi$.
The second element in Eq.~(\ref{eq:hadronicXi}) $\langle\Xi_{c}|\bar{s}\gamma_{\mu}(1-\gamma_{5})b|\Xi_{bc}\rangle$, is formulated as the heavy-to-light transition form factors
\begin{align}
	\langle\Xi_{c}(p_{f})|(V-A)_{\mu}|\Xi_{bc}(p_{i})\rangle = \bar{u}(p_f, s_f) \left[ \gamma_{\mu} f_1(q^2) + i \sigma_{\mu \nu} \frac{q^{\nu}}{m_i} f_2(q^2) + \frac{q_{\mu}}{m_i} f_3(q^2) \right] u(p_i, s_i) \nonumber\\
- \bar{u}(p_f, s_f) \left[ \gamma_{\mu} g_1(q^2) + i \sigma_{\mu \nu} \frac{q^{\nu}}{m_i} g_2(q^2) + \frac{q_{\mu}}{m_i} g_3(q^2) \right] \gamma_5 u(p_i, s_i),\label{eq:transff}
\end{align}
here the momenta and spin of initial and final baryon are denoted by $(p_{i}, s_i)$ and $(p_{f},p_f)$, and the transverse momentum is $q=p_{i}-p_{f}$.

Substituting Eqs.~(\ref{eq:Vdecay}), (\ref{eq:transff}) into Eq.~(\ref{eq:hadronicXi}),
the short-distance amplitude of the weak decay $\Xi_{bc}^{+}\to\Xi_{c}^{+}J/\psi$ can be written as, 
\begin{align}
\mathcal{A}(\Xi_{bc}^{+}\to\Xi_{c}^{+}J/\psi) & =\varepsilon^{\ast\mu}\bar{u}_{\Xi_{c}}\left[A_{SD}^{1}\gamma_{\mu}\gamma_{5}+A_{SD}^{2}\frac{p_{f\mu}}{m_{i}}\gamma_{5}+B_{SD}^{1}\gamma_{\mu}+B_{SD}^{2}\frac{p_{f\mu}}{m_{i}}\right]u_{\Xi_{bc}}.\label{eq:weak amp}
\end{align}
Here the short distance amplitudes $A_{SD}^{1,2},B_{SD}^{1,2}$ are expressed with the following factorization form,
\begin{align}
A_{SD}^{1}&=-\lambda f_{J/\psi}m_{J/\psi} \left ( g_1(q^{2})+g_2(q^{2})\frac{m_{i}-m_{f}}{m_{i}} \right ),
\ \ \ \  A_{SD}^{2}=-2\lambda f_{J/\psi}m_{J/\psi}g_2(q^{2}), \\
B_{SD}^{1}&=\lambda f_{J/\psi} m_{J/\psi} \left ( f_1(q^{2})-f_2(q^{2})\frac{m_{i}+m_{f}}{m_{i}} \right ),
\ \ \ \ \ \ \ \ B_{SD}^{2}=2\lambda f_{J/\psi}m_{J/\psi}f_2(q^{2}),
\end{align}
where the parameter $\lambda$ is $\frac{G_{F}}{\sqrt{2}}V_{CKM}a_{2}(\mu)$ and $m_{i(f)}$ denotes the mass of the initial (final) baryon $\Xi_{bc}(\Xi_{c})$. 
In this work, we take $q^2=m_{J/\psi}^2$ in the calculation of form factors. 

\subsection{Calculation of long-distance contributions using the final states
rescattering mechanism}
\label{subsec:long}

The long-distance contributions are large and very difficult to evaluate.
In this work, we employ the rescattering mechanism of final states
to perform the calculation of the long-distance contributions, as
done in Ref.~\cite{Yu:2017zst,Hu:2024uia}. The rescattering mechanism of final
states can be constructed through the rescattering of two intermediate
particles, as depicted in Fig.~\ref{fig:triangle}. In the following,
we explain the detail of our calculation of the amplitude of the
nonleptonic decay $\Xi_{bc}^{+}\to\Xi_{c}^{+}J/\psi$. At the quark level,
the first weak vertex of the triangle diagram is induced by {the external W emission diagram $T$  to avoid double counting, which is dominated by the factorization contribution.}
The upcoming scattering process can occur
via either an $s$ channel or a $t/u$ channel. The diagram of the $s$
channel would make a sizable contribution when the mass of the exchanged
particle is the sequel to the ones of the mother particle $\Xi_{bc}$.
In bottom quark decays, in contrast to charm quark decays, the resonant FSIs will be expected to be suppressed relative to the rescattering effect arising from quark exchange owing to the lack of the existence of resonances at energies close to the bottom baryon mass. This means that one can neglect the s-channel contribution.
As a result, we only consider the contribution of the $t/u$-channel triangle diagram,
as depicted in Fig.~\ref{fig:triangle}. The remaining two strong interaction vertices, as depicted in Fig.~\ref{fig:triangle}, can be assessed utilizing the hadronic strong interaction Lagrangians~\cite{Aliev:2010yx,Yan:1992gz,Casalbuoni:1996pg,Meissner:1987ge,Li:2012bt,Aliev:2010nh,Xiao:2019mvs,Wu:2021ezz} presented as follows.
The phenomenological effective Lagrangian governing the strong interaction among the doubly heavy baryon, singly heavy baryon, and $D$ mesons is, 
\begin{eqnarray}
&&\mathcal{L}_{P{\mathcal{B}}{\mathcal{B}}} =g_{P{\mathcal{B}}{\mathcal{B}}}Tr[\overline{{\mathcal{B}}}i\gamma_{5}P{\mathcal{B}}],\label{eq:PBB}\\
&&\mathcal{L}_{V{\mathcal{B}}{\mathcal{B}}} =f_{1V{\mathcal{B}}{\mathcal{B}}}Tr[\overline{\mathcal{B}}\gamma_{\mu}V^{\mu}{\mathcal{B}}]+\frac{f_{2V{\mathcal{B}}{\mathcal{B}}}}{m_{{\mathcal{B}}_1}+m_{{\mathcal{B}}_2}}Tr[\overline{\mathcal{B}}\sigma_{\mu\nu}\partial^{\mu}V^{\nu}{\mathcal{B}}],\label{eq:VBB}\\
&&\mathcal{L}_{DDJ/\psi} = ig_{DDJ/\psi} \psi^\alpha \left( \partial_\alpha \bar{D} D - D \partial_\alpha \bar{D} \right), \\
&&\mathcal{L}_{DD^*J/\psi} = -g_{D^*DJ/\psi} \epsilon^{\alpha\beta\rho\tau} \partial_\alpha \psi_\beta \left( \partial_\rho D^*_\tau \bar{D} + D \partial_\rho \bar{D}^*_\tau \right), \\
&&\mathcal{L}_{D^*D^*J/\psi} = ig_{D^*D^*J/\psi} \left[ \psi^\alpha \left( \partial_\alpha D^{*\beta} \bar{D}^*_\beta - D^{*\beta} \partial_\alpha \bar{D}^*_\beta \right) \right. \nonumber\\
&&+ \left( \partial_\alpha \psi^\beta D^*_\beta - \psi^\beta \partial_\alpha D^*_\beta \right) \bar{D}^{*\alpha} + \left. D^{*\alpha} \left( \psi^\beta \partial_\alpha \bar{D}^*_\beta - \partial_\alpha \psi^\beta \bar{D}^*_\beta \right) \right].
\end{eqnarray}
where ${\mathcal{B}}$ denotes the $J^P=\frac{1}{2}^+$ baryon.
Expressions of the vertex elements for all vertices can be derived from above effective Lagrangian directly,
\begin{eqnarray} 
 &  & \langle{\Xi}_{c}(p^{\prime},s_{z}^{\prime})D(q)|{\cal L}_{D\Xi_{c}\Xi_{cc}}|{\Xi}_{cc}(p,s_{z})\rangle=g_{D\Xi_{c}\Xi_{cc}}\bar{u}(p^{\prime},s_{z}^{\prime})i\gamma_{5}u(p,s_{z}),\nonumber\\
 &  & \langle{\Xi}_{c}(p^{\prime},s_{z}^{\prime})D^{*}(q)|{\cal L}_{D^{*}\Xi_{c}\Xi_{cc}}|{\Xi}_{cc}(p,s_{z})\rangle\nonumber \\
 &  & =f_{1D^{*}\Xi_{c}\Xi_{cc}}\bar{u}(p^{\prime},s_{z}^{\prime})\slashed\varepsilon^*(q)u(p,s_{z})+\frac{f_{2D^{*}\Xi_{c}\Xi_{cc}}}{m_{{\Xi_{c}}}+m_{{\Xi_{cc}}}}\bar{u}(p^{\prime},s_{z}^{\prime})\sigma_{\mu\nu}(-iq^{\mu})\varepsilon^{*\nu}(q)u(p,s_{z}),\nonumber \\
&&\left\langle {D(q)J/\psi (p')|\left. {D(p)} \right\rangle } \right. = g_{DDJ/\psi }^D({q^2}){\varepsilon_\alpha ^* (p')}{(p + q)^\alpha },\notag\\\notag	
&&\left\langle {D(q)J/\psi (p')|\left. {{D^*}(p)} \right\rangle } \right. =  - g_{D{D^*}J/\psi }^D({q^2}){\epsilon ^{\alpha \beta \rho \tau }}{\varepsilon^* _\alpha(p')}{\varepsilon_\beta(p)}{p_\rho }p{'_\tau },\\
\notag
&&\left\langle {{D^*}(q)J/\psi (p')|\left. {D(p)} \right\rangle } \right. =  g_{D{D^*}J/\psi }^{{D^*}}({q^2}){\epsilon ^{\alpha \beta \rho \tau }}{\varepsilon^* _\alpha(p')}{\varepsilon _\beta(q)}p{'_\rho }{p_\tau },\\\notag
&&\left\langle {{D^*}(q)J/\psi (p')|\left. {{D^*}(p)} \right\rangle } \right. = g_{{D^*}{D^*}J/\psi }^{{D^*}}[{(p + q)^\alpha }{\varepsilon^{*}_{\alpha }(p') }{\varepsilon ^\beta(p)}\varepsilon_\beta ^{*}(q), \\
&&- {(p + p')^\alpha }{\varepsilon ^{*}_{\alpha }(q)}\varepsilon _\beta ^*(p'){\varepsilon ^\beta(p)} - {(q + p')^\alpha }{\varepsilon _\alpha(p)}{\varepsilon ^{*\beta }(p')}\varepsilon _\beta ^{*}(q)],
\end{eqnarray}
where $\varepsilon$s denoted with momentum are the polarization vectors of $J/\psi$ and $D^{*}$, $\epsilon^{\alpha\beta\rho\tau}$ is the Levi-Civita tensor, and $q=p-p^{\prime}$. The superscripts of $g$ in these equations indicate the off-shell intermediate mesons, while the subscripts refer to the type of vertices. For example, $g_{DDJ/\psi }^{J/\psi }$ represents the strong coupling constant of the vertex $DDJ/\psi$, where $J/\psi$ is the intermediate meson.

There are several methods to calculate the triangle amplitude~\cite{Li:2002pj,Ablikim:2002ep,Li:1996cj,Dai:1999cs,Locher:1993cc,Cheng:2004ru,Lu:2005mx}. The main difference between them is how to deal with the integration of hadronic loops. In Ref.~\cite{Han:2021azw}, only the absorptive (imaginary) part of these diagrams was calculated by using the optical theorem and Cutkosky cutting rule as in Ref.~\cite{Cheng:2004ru}. The real parts of the triangle diagram can be obtained from the dispersion relation in principle. However, due to large uncertainties, it is difficult to give a reliable description of the amplitude over the full region. As a result, this effect is usually neglected in many works. To obtain the full amplitude of the triangle diagram accurately, the most straightforward way is to compute the loop integral directly. In this work, we use the Passarino-Veltman method to reduce the tensor integrals and apply the integration-by-parts method to get expressions in terms of master integrals.

To clarify, we want to determine the amplitude
of the decay mode for ${\Xi}_{bc}(p_{i})\to{\Xi}_{c}(p_{4})J/\psi(p_{3})$. The triangle diagram in Fig.~\ref{fig:triangle}
shows four cases with intermediate states, including $\{D_{s}^{-},{\Xi}_{cc};D_{s}^{-}\}$, $\{D_{s}^{*-},{\Xi}_{cc};D_{s}^{-}\}$, $\{D_{s}^{-},{\Xi}_{cc};D_{s}^{*-}\}$ and $\{D_{s}^{*-},{\Xi}_{cc};D_{s}^{*-}\}$.
The four cases 
involve two kinds of weak vertex ${\Xi}_{bc}(p_{i})\to{\Xi}_{cc}(p_{2})D_{s}(p_{1})$ and ${\Xi}_{bc}(p_{i})\to{\Xi}_{cc}(p_{2})D^{*}_{s}(p_{1})$, corresponding the rescattering amplitudes of ${\Xi}_{cc}(p_{2})D(p_{1})\to{\Xi}_{c}(p_{4})J/\psi(p_{3})$ and ${\Xi}_{cc}(p_{2})D_{s}^{*}(p_{1})\to{\Xi}_{c}(p_{4})J/\psi(p_{3})$, respectively.
The weak vertex can be calculated using the factorization approach as follows, 
\begin{align}
\langle\Xi_{cc}D_{s}^{(*)}|\mathcal{H}_{eff}|\Xi_{bc}\rangle^{C}&=\frac{G_{F}}{\sqrt{2}}V_{cs}^{\ast}V_{bc}a_{1}(\mu)\langle D_{s}^{(*)}|\bar{s}\gamma^{\mu}(1-\gamma_{5})c|0\rangle\langle{\Xi}_{cc}|\bar{c}\gamma_{\mu}(1-\gamma_{5})b|{\Xi}_{bc}\rangle,\label{eq:hadronicbc}
\end{align}
here $a_{1}(\mu)=C_{1}(\mu)+C_{2}(\mu)/3$ is the effective Wilson
coefficients. The first matrix element in Eq.~(\ref{eq:hadronicbc}) can be defined using the decay constant of the emitted meson $D_{s}^{(*)}$, denoted by $f_{D_{s}^{(*)}}$. 
\begin{align}
\langle D_{s}|\bar{s}\gamma^{\mu}(1-\gamma_{5})c|0\rangle & =if_{D_{s}}p^{\mu}_{1},\label{eq:Ddecay}\\
\langle D_{s}^{*}|\bar{s}\gamma^{\mu}(1-\gamma_{5})c|0\rangle & =m_{D_{s}^{*}}f_{D_{s}^{*}}\varepsilon^{\ast\mu}(p_{1}),\label{eq:DSdecay}
\end{align}
where $\epsilon^{\mu}$ is the polarization vector of $D_{s}^{*}$ and $m_{D_{s}^{*}}$ is the mass of the vector meson $D_{s}^{*}$. The second matrix elements can also be effectively parametrized in terms of the form factors $f_{1,2}$ and $g_{1,2}$ in Eq.~(\ref{eq:transff}).
Substituting Eq.~(\ref{eq:transff}) and Eqs.~(\ref{eq:Ddecay}), (\ref{eq:DSdecay}) into Eq.~(\ref{eq:hadronicXi}), the matrix element of the weak vertex $\Xi_{bc}\to\Xi_{cc}D_{s}^{(*)}$ can be written as, 
\begin{align}
	\langle{\Xi}_{cc}D_{s}|\mathcal{H}_{eff}|{\Xi}_{bc}\rangle^{T} & =i\bar{u}_{{\Xi}_{cc}}\left[A_{LD}+B_{LD}\gamma_{5}\right]u_{{\Xi}_{bc}},\label{eq:B2BP}\\
	\langle{\Xi}_{cc}D_{s}^{*}|\mathcal{H}_{eff}|{\Xi}_{bc}\rangle^{T} & =\varepsilon^{\ast\mu}(p_{1})\bar{u}_{{\Xi}_{cc}}\left[A_{LD}^{1}\gamma_{\mu}\gamma_{5}+A_{LD}^{2}\frac{p_{2\mu}}{m_{i}}\gamma_{5}+B_{LD}^{1}\gamma_{\mu}+B_{LD}^{2}\frac{p_{2\mu}}{m_{i}}\right]u_{{\Xi}_{bc}}.\label{eq:B2BV}
\end{align}
Here the long-distance amplitudes $A_{LD},B_{LD}$ and $A_{LD}^{1,2},B_{LD}^{1,2}$ are specifically designated to encapsulate the strong interaction information via factorization,
\begin{align}
A_{LD}&=\lambda f_{D_{s}}(m_{i}-m_{2})f_1(q^{2}), \hspace{3cm}   B_{LD}=\lambda f_{D_{s}}(m_{i}+m_{2})g_1(q^{2}),\\
A_{LD}^{1}&=-\lambda f_{D_{s}^{*}}m_{D_{s}^{*}} \left ( g_1(q^{2})+g_2(q^{2})\frac{m_{i}-m_{2}}{m_{i}} \right ),
\ \ \ \  A_{LD}^{2}=-2\lambda f_{D_{s}^{*}}m_{D_{s}^{*}}g_2(q^{2}), \\
B_{LD}^{1}&=\lambda f_{D_{s}^{*}}m_{D_{s}^{*}} \left ( f_1(q^{2})-f_2(q^{2})\frac{m_{i}+m_{2}}{m_{i}} \right ),
\ \ \ \ \ \ \ \ B_{LD}^{2}=2\lambda f_{D_{s}^{*}}m_{D_{s}^{*}}f_2(q^{2}),
\end{align}
where the parameter $\lambda$ is $\frac{G_{F}}{\sqrt{2}}V_{CKM}a_{1}(\mu)$.
Here we take $q^2=m_{D_{s}^{(*)}}^2$ in the calculation of the long-distance contribution.
$m_{2}$ denotes the mass of intermediate state $\Xi_{cc}$, while the rescattering amplitude is computed using
the hadronic effective Lagrangian. We use the symbol $\mathcal{M}[P_{1},P_{2};P_{k}]$ to represent a
triangle amplitude in this work. The amplitudes of the diagrams shown in Fig.~\ref{fig:triangle}
can be expressed as follows, 

\begin{align}
M[D_s^-,\Xi_{cc}^{++};D_s^-]
&= -i\!\! \int \frac{d^4p_{1}}{(2\pi)^4}(2\pi)^4\delta^4(p_3+p_4-p_1)g_{\psi DD}^Dg_{{\Xi_{cc} \Xi_c}D_s}{\cal {PF}}\nonumber\\
&\quad \times \varepsilon^{\alpha}(p_3)(2p_{1\alpha}-p_{3\alpha})\bar{u}_{\Xi_c}\gamma_5({\slashed p}_{2}+m_{2})\left[A_{LD}+B_{LD}\gamma_{5}\right]u_{\Xi_{bc}},
\end{align}
\begin{align}
M[D_s^{*-},\Xi_{cc}^{++};D_s^-]
&= \!\! \int \frac{d^4p_{\text{\tiny $1$}}}{(2\pi)^4}(2\pi)^4\delta^4(p_3+p_4-p_1)g_{D^*DJ/\psi}^Dg_{\Xi_{cc}\Xi_cD_s}{\cal {PF}}\nonumber\\
&\quad \times \epsilon^{\alpha\beta\rho\tau}\varepsilon_{\alpha}(p_{3})\varepsilon_{\beta}(p_{1})p_{1\rho}p_{3\tau} \bar{u}_{\Xi_c}\gamma_5({\slashed p}_{2}+m_{2})\nonumber\\
&\quad \times \varepsilon^{*\mu}(p_{1})\left[A_{LD}^{1}\gamma_{\mu}\gamma_{5}+A_{LD}^{2}\frac{p_{2\mu}}{m_{i}}\gamma_{5}+B_{LD}^{1}\gamma_{\mu}+B_{LD}^{2}\frac{p_{2\mu}}{m_{i}}\right]u_{\Xi_{bc}},
\end{align}
\begin{align}
M[D_s^{-},\Xi_{cc}^{++};D_s^{*-}]
&=- \!\! \int \frac{d^4p_{\text{\tiny $1$}}}{(2\pi)^4}(2\pi)^4\delta^4(p_3+p_4-p_1)g_{DDJ/\psi}^{D^{*}}{\cal {PF}}\nonumber\\
&\quad \times \epsilon^{\alpha\beta\rho\tau}\varepsilon_{\alpha}(p_{3})\varepsilon_{\beta}(k)p_{3\rho}p_{1\tau}
\bar{u}_{\Xi_c}(f_{1VBB}\gamma_{\nu}-\frac{if_{2VBB}}{m_2+m_4}\sigma_{\mu\nu}k^{\mu})\nonumber\\&\quad \times \varepsilon^{*\nu}(k)({\slashed p}_{2}+m_{2}) \left[A_{LD}+B_{LD}\gamma_{5}\right]u_{\Xi_{bc}},
\end{align}
\begin{align}
M[D_s^{*-},\Xi_{cc}^{++};D_s^{*-}]
&= i\!\! \int \frac{d^4p_{\text{\tiny $1$}}}{(2\pi)^4}(2\pi)^4\delta^4(p_3+p_4-p_1)g_{D^*D^*J/\psi}^{D^*}{\cal {PF}}\nonumber\\
&\quad \times[(p_3+k)^{\alpha}\varepsilon^*_{\alpha}(p_{3})\varepsilon_{\beta}(p_1)\varepsilon^*_{\beta}(k)-(p_1+p_{3})^{\alpha}\varepsilon^*_{\alpha}(k)\varepsilon_{\beta}^{*}(p_{3})\varepsilon^{\beta}(p_{1})\nonumber\\
&\quad \quad -(k+p_3)^{\alpha}\varepsilon_{\alpha}(p_{1})\varepsilon^*_\beta(p_{3})\varepsilon^{*\beta}(k)]\nonumber\\
&\quad \times\bar{u}_{\Xi_c}(f_{1VBB}\gamma_{\nu}-\frac{if_{2VBB}}{m_2+m_4}\sigma_{\rho\nu}k^{\rho})\varepsilon^{\nu}(k)({\slashed p}_{2}+m_{2})\nonumber\\
&\quad \times \varepsilon^{*\mu}(p_{1})\big[A_{LD}^{1}\gamma_{\mu}\gamma_{5}+A_{LD}^{2}\frac{p_{2\mu}}{m_{i}}\gamma_{5}+B_{LD}^{1}\gamma_{\mu}+B_{LD}^{2}\frac{p_{2\mu}}{m_{i}}\big]u_{\Xi_{bc}},
\end{align}

where the note ${\cal {PF}}$ represents the multiply of the propagators and form factor, 
\begin{eqnarray}
{\cal {PF}} & = & \frac{1}{(p_{1}^{2}-m_{1}^{2}+i\epsilon)(p_{2}^{2}-m_{2}^{2}+i\epsilon)(p_{k}^{2}-m_{k}^{2}+i\epsilon)}\left(\frac{\Lambda_{k}^{2}-m_{k}^{2}}{\Lambda_{k}^{2}-p_{k}^{2}}\right).\label{eq:TF}
\end{eqnarray}
In the given equation, the strong coupling constants $g_{DDJ/\psi}$, $f_{1{\cal {B}}_{c}{\cal {B}}_{c}V}$ and $f_{2{\cal {B}}_{c}{\cal {B}}_{c}V}$ are calculated in the on-shell renormalization scheme. The exchange states $D^{(*)}_{s}$ are generally off shell, which makes the extracted values of the strong coupling constants unreliable. The form factors of the exchange particles $F(p_{j},m_{j})$ are introduced to take into account the off-shell effects as well as to make the theoretical framework self-consistent~\cite{Cheng:2004ru}. Furthermore, it is necessary to apply an appropriate regularization scheme to regulate the divergence inevitably appearing in the master integrals of the triangle diagram amplitudes. The form factors introduced in Eq.\eqref{eq:TF} correspond to the Pauli-Villars regularization scheme
\begin{align}
F(p_{k},m_{k})=\left(\frac{\Lambda_{k}^{2}-m_{k}^{2}}{\Lambda_{k}^{2}-p_{k}^{2}}\right)^{n}.\label{eq:Ffactor}
\end{align}
The cutoff $\Lambda_{k}$ is given by $\Lambda_{k}=m_{k}+\eta\Lambda_{{\rm QCD}}$, with $\Lambda_{{\rm QCD}}=220\text{MeV}$~\cite{Cheng:2004ru}. The phenomenological parameter $\eta$ is determined from experimental data, which can not be derived from the first-principle method. In this work we will use the experimental data of the similar channel ${\Lambda}_{b}\to{\Lambda}J/\psi$ to determine the value of the parameter $\eta$. The form factors given in Eq.~(\ref{eq:Ffactor}) usually show monopole or dipole behavior, where the exponential factor $n$ takes values 1 or 2. The branching ratios for $B$-meson decays are similar for both choices in Ref.~\cite{Cheng:2004ru}, and then we choose $n=1$ for simplicity.
		
After gathering all the fragments, the amplitude of decay ${\Xi}_{bc}^{+}\to{\Xi}_{c}^{+}J/\psi$
can be expressed as: 
\begin{align}
\mathcal{A}({\Xi}_{bc}^{+}\to{\Xi}_{c}^{+}J/\psi) & =\mathcal{M}_{SD}({\Xi}_{bc}^{+}\to{\Xi}_{c}^{+}J/\psi)+\mathcal{M}[D_{s}^{-},\Xi_{cc}^{++};D_{s}^{-}]+\mathcal{M}[D_{s}^{*-},\Xi_{cc}^{++};D_{s}^{-}]\nonumber \\
&+\mathcal{M}[D_{s}^{-},\Xi_{cc}^{++};D_{s}^{*-}] +\mathcal{M}[D_{s}^{*-},\Xi_{cc}^{++};D_{s}^{*-}],
\end{align}
where $\mathcal{M}_{SD}$ labels its short-distance contributions {including the factorization contributions of $C$ diagrams}. 

\subsection{Observables}
To clarify the several observables of process ${\Xi}_{bc}^{+}\to{\Xi}_{c}^{+}J/\psi$, it started with the spin of ${\Xi_{bc}^+}$ which has not yet been measured but the quark model prediction is $1/2$. This mode is therefore the decay of a spin $1/2$ particle into a spin $1$ and a spin $1/2$ particle. In the helicity formalism, the decay can be described by four helicity amplitudes, 
\begin{align}
    H_{\lambda_{f},\lambda_{V}}^{\lambda_{i}} \sim \left\{H_{-{1\over 2},0}^{{1\over2}}, H_{{1\over 2},0}^{-{1\over 2}}, H_{-{1\over 2},-1}^{-{1\over 2}}, H_{{1\over 2},1}^{{1\over 2}}\right\},
\end{align}
where $\lambda_{i},\lambda_{f},\lambda_{V}$ are the helicity of ${\Xi_{bc}}$, ${\Xi_{c}}$ and $J/\psi$, respectively. In this decay process, the helicity amplitudes can be obtained as follows, 
\begin{eqnarray}
 & HV_{\lambda_{f},\lambda_{V}}^{\lambda_{i}} & =\langle{\cal B}_{f}(\lambda_{f})|\bar{c}\gamma^{\mu}b|{\cal B}_{i}(\lambda_{i})\rangle\varepsilon_{\mu}^{*}(\lambda_{V}),\nonumber \\
 & HA_{\lambda_{f},\lambda_{V}}^{\lambda_{i}} & =\langle{\cal B}_{f}(\lambda_{f})|\bar{c}\gamma^{\mu}\gamma_{5}b|{\cal B}_{i}(\lambda_{i})\rangle\varepsilon_{\mu}^{*}(\lambda_{V}),\nonumber \\
 & H_{\lambda_{f},\lambda_{V}}^{\lambda_{i}}= & HV_{\lambda_{f},\lambda_{V}}^{\lambda_{i}}-HA_{\lambda_{f},\lambda_{V}}^{\lambda_{i}}.\label{eq:helicity}
\end{eqnarray}
The helicity amplitudes are related to the transition form factors as follows:
\begin{eqnarray}
HV_{-\frac{1}{2},0}^{\frac{1}{2}} & = & HV_{\frac{1}{2},0}^{-\frac{1}{2}}=-i\sqrt{\frac{Q_{-}}{q^{2}}}\big((m_{i}+m_{f})f_{1}-\frac{q^2}{m_{i}}f_{2}\big),\nonumber \\
HV_{\frac{1}{2},1}^{\frac{1}{2}} & = & HV_{-\frac{1}{2},-1}^{-\frac{1}{2}}=i\sqrt{2Q_{-}}(-f_{1}+\frac{m_{i}+m_{f}}{m_{i}}f_{2}),\\
HA_{-\frac{1}{2},0}^{\frac{1}{2}} & = & -HA_{\frac{1}{2},0}^{-\frac{1}{2}}=-i\sqrt{\frac{Q_{+}}{q^{2}}}((m_{i}-m_{f})g_{1}+\frac{q^2}{m_{i}}g_{2}),\nonumber \\
HA_{\frac{1}{2},1}^{\frac{1}{2}} & = & -HA_{-\frac{1}{2},-1}^{-\frac{1}{2}}=i\sqrt{2Q_{+}}(-g_{1}-\frac{m_{i}-m_{f}}{m_{i}}g_{2}).
\end{eqnarray}
Then the decay width can be given as follows,
\begin{align}
  \Gamma({\Xi_{bc}}\to{\Xi_{c}}J/\psi)
  =\frac{|\vec{p}_{c}|}{8\pi m_i^2}\frac{1}{2}\sum_{\lambda_{i},\lambda_{f},\lambda_{V}}|H_{\lambda_{f},\lambda_{V}}^{\lambda_{i}}({\Xi_{bc}}\to{\Xi_{c}}J/\psi)|^2,\label{eq:widthc}
  \end{align}
here $|\vec{p}_{c}|$ is the magnitude of the three-momentum of ${\Xi_{c}}$
in the rest frame of ${\Xi_{bc}}$.

\section{Numerical results and discussions}
\label{sec:results}
\subsection{Input parameters}
Calculation of this work requires inputs such as initial and final state masses, decay constants of $D$ meson and charmonium $J/\psi$, strong
couplings, and the
lifetimes of the bottom-charmed baryons.
The mass and lifetime of bottom-charmed baryon have been studied in many theoretical works~\cite{Workman:2022ynf,Chen:2016spr,Yu:2018com,Yu:2019lxw,Cheng:2018mwu,Berezhnoy:2018bde}. 
 The results from Refs.\cite{Brown:2014ena,Berezhnoy:2018bde}
are adopted in this work,
$m_{\Xi_{bc}^{+}}=6.943$ MeV, $\tau_{\Xi_{bc}^{+}}=0.24\pm0.02$~ps.
The LHCb collaboration has successfully measured the mass of $\Xi_{cc}^{++}$: $m_{\Xi_{cc}^{++}}=3.621$ MeV\cite{Aaij:2017ueg}. The masses of singly charmed baryons $\Xi_{c}$,
$D$ mesons and charmonia $J/\psi$, can be easily found in Particle
Data Group~\cite{ParticleDataGroup:2024cfk}. 
The decay constants of $D$ mesons and charmonia $J/\psi$ are obtained from the literature~\cite{ParticleDataGroup:2024cfk,Choi:2015ywa,Feldmann:1998vh}, and summarized in Tab.~\ref{table:decayconstants}.
In addition, the nonperturbative strong coupling constants serve a pivotal role in our calculation. 
Most of them can be taken from theoretical works. In this work, we use theoretical results~\cite{OsorioRodrigues:2013xvc,OsorioRodrigues:2015rrh,Khosravi:2013ad,Hu:2025ajx,Khodjamirian:2011jp}. In order to determine the model parameter $\eta$, the input parameters related to the decay $\Lambda_{b}\to\Lambda J/\psi$ are also needed.
In Tab.~\ref{table:decayconstants}, we list the input parameter values used in this work. Among them, the strong couplings $g_{\Lambda_{c}^{+}\Lambda^{0}D_{s}^{*+}}$ and $g_{\Lambda_{c}^{+}\Lambda^{0}D_{s}}$ are determined through $SU(3)$ flavor symmetry using the results of $g_{\Lambda_{c}^{+}ND^{(*)}}$ from the LCSR work~\cite{Khodjamirian:2011jp}. 

The transition form factors of $\Xi_{bc}\to\Xi_{c}$, $\Xi_{bc}\to\Xi_{cc}$ and $\Lambda_{b}\to\Lambda_{c}$, $\Lambda_{b}\to\Lambda$ are obtained from theoretical works~\cite{Hu:2020mxk,Wang:2017mqp,Ke:2019smy,Wei:2009np} within the light-front quark model. 
For the convenience of the paper, we do not list them in this paper, readers can find them in the references.

\begin{table}[H]
\centering \caption{{Decay constants (in units of MeV) of the light mesons~\cite{ParticleDataGroup:2024cfk,Choi:2015ywa} and strong couplings~\cite{Cheng:2004ru,Nakayama:2006ps,Aliev:2010yx,Aliev:2010nh,Aliev:2010ev,Yan:1992gz,Casalbuoni:1996pg,Meissner:1987ge,Li:2012bt} used in this work.}}
\label{table:decayconstants} %
\begin{tabular}{cccc}
\hline\hline 
$f_{D_{s}}$~  & ~$f_{D_{s}^{*}}$~  & ~$f_{J/\psi}$~  \\\hline 
$249.9\pm0.5$~\cite{ParticleDataGroup:2024cfk}  &$314^{+19}_{-14}$~\cite{Gelhausen:2013wia}& $407\pm5$~\cite{ParticleDataGroup:2014cgo}  \\\hline
$g_{D_{s}D_{s}J/\psi}$~  &~$g_{D_{s}^{*}D_{s}J/\psi}$~   & ~$g_{D_{s}^{*}D_{s}^{*}J/\psi}$~  \\\hline
$6.24^{+0.41}_{-0.36}$~\cite{OsorioRodrigues:2013xvc}& $4.39^{+1.44}_{-1.30}$~\cite{OsorioRodrigues:2015rrh}  & $4.96\pm1.42$~\cite{Khosravi:2013ad}\\\hline
~$g_{\Xi_{cc}^{++}\Xi_{c}^{+}D_{s}^{+}}$~  & ~$g_{\Xi_{cc}^{++}\Xi_{c}^{+}D_{s}^{*+}}$~&$g_{\Lambda_{c}^{+}\Lambda^{0}D_{s}^{+}}$ ~&$g_{\Lambda_{c}^{+}\Lambda^{0}D_{s}^{*+}}$\\\hline
 $2.17\pm0.68$~\cite{Hu:2025ajx}  & $\{2.44\pm0.83,6.68\pm2.38\}$~\cite{Hu:2025ajx} & $-8.74^{+3.51}_{-4.32}$~\cite{Khodjamirian:2011jp} &\{$4.47^{+2.04}_{-1.71}$,$-2.94^{+1.47}_{2.37}$\}~\cite{Khodjamirian:2011jp}\\\hline\hline
\end{tabular}
\end{table}

\subsection{The model parameter $\eta$}
The model parameter $\eta$ cannot be obtained from the first-principle but can be determined from experimental data. Due to limitations in the experimental data for  bottom-charmed baryon decays, we extracted $\eta$ value from the analogous decay of a singly bottom baryon via the same quark-level channel. Similarly,  the triangle diagrams for $\Lambda_{b}\to\Lambda J/\psi$ can also be calculated using the factorization approach and final state interactions theory as follows: 
 \begin{align}
		\mathcal{A}({\Lambda}_{b}\to{\Lambda}J/\psi) & =\mathcal{M}_{SD}({\Lambda}_{b}\to{\Lambda}J/\psi)+\mathcal{M}[D_{s}^{-},\Lambda_{c}^{+};D_{s}^{-}]+\mathcal{M}[D_{s}^{*-},\Lambda_{c}^{+};D_{s}^{-}]\nonumber \\
		&+\mathcal{M}[D_{s}^{-},\Lambda_{c}^{+};D_{s}^{*-}] +\mathcal{M}[D_{s}^{*-},\Lambda_{c}^{+};D_{s}^{*-}].
		\end{align}

The branching ratio of $\Lambda_{b}\to\Lambda J/\psi$ can then be calculated. The data in Eq.~(\ref{eq:bbb}) is mainly from the D0 experiment at the Tevatron proton-proton collider with a center-of-mass energy of $1.96$ TeV. The fragmentation fraction $f_{\Lambda_{b}}$ depends on experimental conditions. For example, the LHCb experiment is performed at center-of-mass energies of $7$, $8$, and $13$ TeV, so its fragmentation fractions cannot be used here. We now briefly discuss the value of $f_{\Lambda_{b}}$. The total fragmentation fraction of b baryons is the sum of all weakly decaying b baryons,
\begin{eqnarray}
f_\text{baryon}=f_{\Lambda_{b}}\left(1+2\frac{f_{\Xi_{b}}}{f_{\Lambda_{b}}}+\frac{f_{\Omega_{b}}}{f_{\Lambda_{b}}}\right).
\end{eqnarray}
Since the production of $\Omega_{b}$ is suppressed relative to $\Xi_{b}$ due to an additional strange quark, $f_{\Omega_{b}}/f_{\Lambda_{b}}$ can be neglected. 

Hence, the ratio $f_{\Xi_{b}}/f_{\Lambda_{b}}$ plays a critical role in elucidating $\Lambda_{b}$ production mechanisms. According to the LHCb Collaboration analysis~\cite{LHCb:2019sxa}, the fragmentation fraction ratios were determined as ${f_{\Xi_{b}^{-}}}/{f_{\Lambda_{b}^{0}}}=(6.7\pm0.5\pm0.5\pm2.0)\%$ at $\sqrt{s}=7,8~\text{TeV}$, and $(8.2\pm0.7\pm0.6\pm2.5)\%$ at $\sqrt{s}=13~\text{TeV}$. As analyzed in Ref.~\cite{Jiang:2018iqa}, the corresponding ratio ${f_{\Xi_{b}^{0}}}/{f_{\Lambda_{b}^{0}}}$ equals $(5.4\pm2.0)\%$. These three analyses demonstrate mutual consistency while remaining significantly below unity, with minor discrepancies deemed negligible. Consequently, we adopt $f_{\Xi_{b}^{0}}/f_{\Lambda_{b}^{0}}=(5.4\pm2.0)\%$ for subsequent calculations of $f_{\Lambda_{b}}$. The average total baryon production fraction is derived from HFLAV collaboration data~\cite{HFLAV:2016hnz}, 
\begin{eqnarray}
f_\text{baryon} = 0.196\pm0.046,\quad \text{at Tevatron}.
\end{eqnarray}
This result corresponds to $f_{\Lambda_b} = 0.177 \pm 0.042$, leading to a branching ratio of $\text{Br}(\Lambda_b \to \Lambda J/\psi) = (3.3 \pm 0.9) \times 10^{-4}$.

\begin{figure}[ht]
\centering
\includegraphics[width=0.80\textwidth]{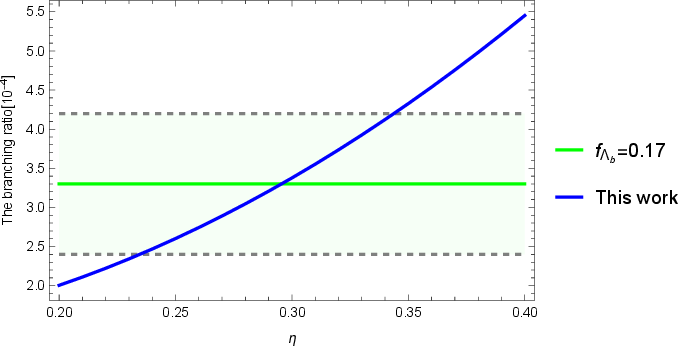}
	\caption{The value of $\eta$ determined by the branching ratio of the decay $\Lambda_{b} \to\Lambda J/\psi$.}
	\label{fig:etavalue}
\end{figure}
The determined range restricts $\eta$ to $0.3\pm0.05$, as illustrated in Fig.~\ref{fig:etavalue}. The branching ratio for the nonleptonic decay $\Lambda_{b}\to\Lambda J/\psi$ is calculated by incorporating $\eta=0.3\pm0.05$ to propagate the parameter uncertainty, combined with additional inputs such as lifetimes and masses of the singly bottomed baryons~\cite{ParticleDataGroup:2024cfk} summarized in Table~\ref{tab:resultforlambda}. This table additionally compares the present results with those from alternative theoretical investigations of the $\Lambda_{b}\to\Lambda J/\psi$ decay process.

\begin{table}[!htbh]
\scriptsize
\caption{
Theoretical results on the branching ratios ($10^{-4}$) for the $\Lambda_b\rightarrow \Lambda J/\psi$ decays, where the theoretical errors are from the model parameter $\eta$.
For comparison, 
the available predictions from perturbative QCD(pQCD)~\cite{Rui:2022sdc,Chou:2001bn}, covariant confined quark model(CCQM)~\cite{Gutsche:2018utw}, nonrelativistic quark model(NRQM)~\cite{Cheng:1996cs,Mott:2011cx}, relativistic three quark model(RTQM)~\cite{Ivanov:1997ra}, the effective Lagrangian with factorization framework(EL with FF)~\cite{Fayyazuddin:2017sxq}, covariant oscillator quark model(COQM)~\cite{Mohanta:1998iu}, quark model(QM)~\cite{Cheng:1995fe}, factorization approach(FA)~\cite{Hsiao:2015cda}, and LFQM~\cite{Wei:2009np} are also presented.
}
\label{tab:resultforlambda}
\begin{tabular}{c|ccccccc}
\hline\hline
 \multirow{4}{*}{$\mathcal{B}(\Lambda_b\rightarrow \Lambda J/\psi)$}& This work & pQCD~\cite{Rui:2022sdc} & CCQM~\cite{Gutsche:2018utw} 
 &NRQM~\cite{Cheng:1996cs}&
 RTQM~\cite{Ivanov:1997ra}&
  EL with FF~\cite{Fayyazuddin:2017sxq}
 \\\cline{2-7}
 & $3.38_{-0.77}^{+0.95}$ & $7.75_{-3.17-0.96}^{+4.08+1.39}$ & $8.0$ 
 &1.6&
2.7&7.8\\\cline{2-7}
& COQM~\cite{Mohanta:1998iu} & QM~\cite{Cheng:1995fe} & pQCD~\cite{Chou:2001bn} 
 &FA~\cite{Hsiao:2015cda}&
 LFQM~\cite{Wei:2009np}&
  NRQM~\cite{Mott:2011cx}\\\cline{2-7}
  &2.5
  &2.1
  &3.5 $\pm$ 1.8& 3.3$\pm$2.0
  &8.4
  &8.2 \\
\hline\hline
\end{tabular}
\end{table}

A comparative analysis demonstrates that our theoretical results align with existing findings within their reported uncertainties. This enables the prediction of branching ratios for charmed-bottom baryons by employing $\eta = 0.3 \pm 0.05$.

\subsection{The branching ratio }
Using the parameter $\eta=0.3\pm0.05$ determined by the decay $\Lambda_{b}\to\Lambda J/\psi$ and other input parameters, the branching ratio 
for the decay $\Xi_{bc}^{+}\to\Xi_{c}^{+}J/\psi$ can be predicted as following,

\begin{equation}
  {\mathcal{B}}(\Xi_{bc}^{+}\to\Xi_{c}^{+}J/\psi)=(1.55_{-0.42}^{+0.50}) \times 10^{-4}.\label{eq:branxibc}
\end{equation}

\begin{figure}[h]
    \centering
    \begin{minipage}{0.45\linewidth}
    \centering
    \subfloat[][]{
        \includegraphics[scale=0.7]{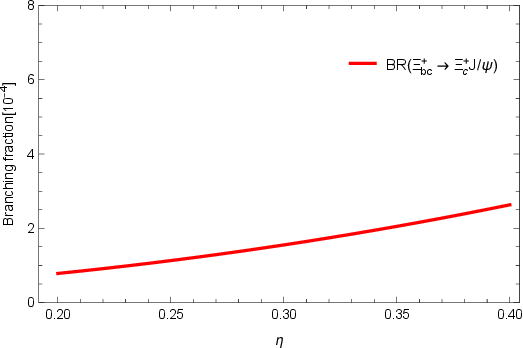}}
    \end{minipage}
    \begin{minipage}{0.45\linewidth}
    \centering
        \subfloat[][]{\includegraphics[scale=0.7]{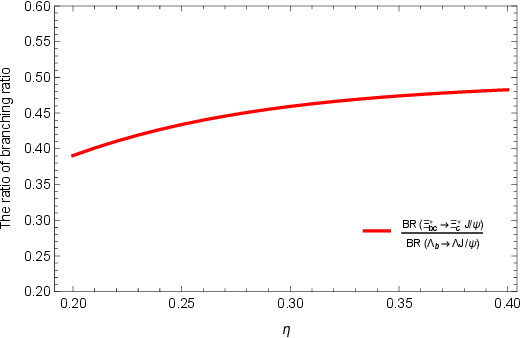}}
    \end{minipage}
\caption{(a):The branching ratio 
of the decay $\Xi_{bc}^{+}\to\Xi_{c}^{+}J/\psi$;
(b):The dependence of the ratio ${\cal B}(\Xi_{bc}^{+}\to\Xi_{c}^{+}J/\psi)/{\cal B}(\Lambda_{b}\to\Lambda J/\psi)$ on the parameter $\eta$.}
	\label{fig:branchanddecay}
\end{figure}

From Eq.~(\ref{eq:branxibc}), the branching ratio for the decay $\Xi_{bc}^{+}\to\Xi_{c}^{+}J/\psi$ is estimated to be in the range $(1.55_{-0.42}^{+0.50})\times10^{-4}$. The uncertainty of the branching ratio is estimated by varying the model parameter $\eta$, since the uncertainties associated with $\eta$ constitute the dominant source of error. Other input parameters introduce smaller uncertainties than the uncertainty from the model parameters, and are thus not included in our error estimation. This value is approximately half of the corresponding branching ratio for the decay $\Lambda_{b}\to\Lambda J/\psi$, as illustrated in Fig.~\ref{fig:branchanddecay}(b), this ratio has the extremely weak dependence on $\eta$. Our numerical results of Fig.~\ref{fig:branchanddecay} are presented in the Supplemental Material \cite{OpenData}. Compared to the decay $B_{c}^{+}\to D_{s}^{+}J/\psi$, the $\Xi_{bc}^{+}\to\Xi_{c}^{+}J/\psi$ process is color-suppressed, resulting in a branching ratio that is one order of magnitude smaller. In the absence of experimental data for the $B_{c}^{+}$ meson decay, we estimate its branching ratio using the ratio ${\cal B}(B_{c}^{+}\to D_{s}^{+}J/\psi)/{\cal B}(B_{c}^{+}\to \pi^{+}J/\psi)=2.76(2.90)$ from the ATLAS (LHCb) collaboration report~\cite{ATLAS:2022aiy,LHCb:2013kwl} and the theoretical prediction of ${\cal B}(B_{c}^{+}\to \pi^{+}J/\psi)=1.97\times10^{-3}$~\cite{Zhang:2023ypl} which is also predicted using the light front quark model, yielding ${\cal B}(B_{c}^{+}\to D_{s}^{+}J/\psi)=5.44(5.71)\times10^{-3}$. Therefore, our theoretical prediction for ${\cal B}(\Xi_{bc}^{+}\to\Xi_{c}^{+}J/\psi)=(1.55_{-0.42}^{+0.50})\times 10^{-4}$ is considered reliable. 
\subsection{The signal events at the LHC}
Bottom-charmed baryons $\Xi_{bc}^{+}, \Xi_{bc}^{0}, \Omega_{bc}^{0}$ contain a bottom quark and a charm quark, which implies the simultaneous production of a $b\bar{b}$ pair and a $c\bar{c}$ pair. Subsequently, one bottom quark and one charm quark, together with a light quark, form the bottom-charmed baryon. Consequently, their production cross-sections are smaller compared to those of doubly charmed baryons.

The analysis in Ref.~\cite{Qin:2021zqx} highlights that the inclusive $\Xi_{bc}\to \Xi_{cc} X$ decay constitutes a golden channel for the experimental discovery of $\Xi_{bc}$ at the LHC. Although smaller by approximately two orders of magnitude compared to the inclusive decay $\Xi_{bc}\to \Xi_{cc} X$, this special channel $\Xi_{bc}^{+}\to\Xi_{c}^{+}J/\psi$ provides a distinct experimental signature characterized by high signal purity through effective background suppression, facilitates complete reconstruction of event kinematics for identification and measurement of intermediate particles, and permits direct comparison with theoretical frameworks.

At LHCb Run III, an integrated luminosity of $23~\rm{fb}^{-1}$ is anticipated~\cite{Qin:2021zqx}, with a projected $\Xi_{bc}^{+}$ production cross-section of $10\sim 25~\text{nb}$ in the LHCb detector~\cite{Berezhnoy:2018bde}. The corresponding production events of $\Xi_{bc}^{+}$ $N_{p}(\Xi_{bc})$ is approximately $4\times 10^8$. Considering the subsequent decay chain involving $Br(\Xi_{c}^{+}\to p K^{-}\pi^{+})=6.2\times10^{-3}$~\cite{Belle:2019bgi,LHCb:2020gge} and $Br(J/\psi\to\mu^{+}\mu^{-})=5.96\times10^{-2}$~\cite{ParticleDataGroup:2024cfk}, combined with detection efficiencies for final states $\epsilon_{p,K^{-},\pi^{+}}=90\sim95\%$~\cite{LHCbRICHGroup:2012mgd,LHCb:2014nio}  and $\epsilon_{\mu^{+},\mu^{-}}=98\sim99\%$~\cite{Archilli:2013npa}, the expected signal events $N_s$ for this decay channel can be estimated as follows, 
\begin{equation}
    N_{s}=N_{p}(\Xi_{bc})Br(\Xi_{bc}^{+}\to\Xi_{c}^{+}J/\psi)Br(\Xi_{c}^{+}\to p K^{-}\pi^{+})Br(J/\psi\to\mu^{+}\mu^{-})(90\%)^3\sim16.
\end{equation}
This result is consistent with the analysis by the LHCb collaboration~\cite{LHCb:2022fbu,LHCb:2020ayi}; specifically, based on $1100$ $B_{c}^{+}\to D_{s}^{+}J/\psi$ candidates observed in the full LHCb dataset, approximately $15$ reconstructed signal decays of $\Xi_{bc}^{+}\to\Xi_{c}^{+}J/\psi$ are expected within the detector acceptance.
While the LHC is designed for proton-proton collisions at a centre-of-mass energy of $14\,\rm{TeV}$ with a peak luminosity of $10^{34} \rm{cm}^{-2} \rm{s}^{-1}$~\cite{Evans:2008zzb}. Then the integrated luminosity will be changed to $100 fb^{-1}/yr$, one can estimate that about $3.5\times10^9$ events per year can be produced~\cite{Zhang:2011hi}. Based on these parameters, the expected signal yield $N_s$ will be $140$.
From the above, we can found that with larger datasets anticipated from future LHCb experiments and the inclusion of additional decay modes, there is potential to provide evidence or achieve a discovery of the baryon $\Xi_{bc}^{+}$.

\section{Summary}
In this study, we investigate the nonleptonic weak decay $\Xi_{bc}^{+} \to \Xi_{c}^{+} J/\psi$ using the final-state interaction approach. The model parameter $\eta$ is constrained to be $0.3\pm0.05$ based on the analogous decay $\Lambda_b \to \Lambda J/\psi$. Using the input parameters such as initial and final state masses, decay constants of $D$ meson and so on, we predict the branching ratio
for $\Xi_{bc}^{+} \to \Xi_{c}^{+} J/\psi$, as given in Eq.~(\ref{eq:branxibc}).
Since this decay is color-suppressed, the ratio $\mathcal{B}(\Xi_{bc}^{+} \to \Xi_{c}^{+} J/\psi)/\mathcal{B}(B_{c}^{+} \to D_{s}^{+} J/\psi)$ should scale as $1/N_c^2$. Our theoretical prediction for $\mathcal{B}(\Xi_{bc}^{+} \to \Xi_{c}^{+} J/\psi)$, $(1.55_{-0.42}^{+0.50})\times 10^{-4}$, is consistent with this scaling argument based on the calculated branching fraction $\mathcal{B}(B_c^{+} \to D_s^{+} J/\psi) \sim  10^{-3}$. This consistency demonstrates the reliability of our decay parameter predictions, which can serve as benchmarks for future experimental investigations. Considering the $\Xi_{bc}$ production rate and detection efficiency for final states containing $p, K^{-}, \pi^{+}$, and $\mu^{+}\mu^{-}$, we anticipate that about $16(140)$ signal events will be collected by the LHCb collaboration when the integrated luminosity of LHC reach to $23(100)\ \rm{fb}^{-1}/\rm{year}$. With increased data samples in forthcoming runs and additional decay channels incorporated, the observation or even discovery of the $\Xi_{bc}^{+}$ baryon becomes highly probable.

\begin{acknowledgments}
The authors are very grateful to Ji-Bo He and Zhi-Peng Xing for helpful discussions on the LHC experimental search and the theoretical framework. This work is supported in part by the National Natural Science Foundation of China under Grants No. 12505114, No. 12247101, No. 12005294, and No. 12335003, the Fundamental Research Funds for the Central Universities under Grants No. lzujbky-2024-oy02 and No. lzujbky-2025-jdzx07, the Natural Science Foundation of Gansu Province under Grant No.25JRRA799, and the ‘111 Center’ under Grant No. B20063.
\end{acknowledgments}

\end{document}